\PassOptionsToPackage{prologue,table,xcdraw,dvipsnames}{xcolor}
\documentclass[sigconf]{acmart}
\AtBeginDocument{%
  }
\usepackage{multirow, booktabs, subcaption}
\usepackage{pifont} 
\usepackage[table,xcdraw,dvipsnames]{xcolor}
\copyrightyear{2025}
\acmYear{2025}
\acmConference[RecSys '25]{Proceedings of the Nineteenth ACM Conference on Recommender Systems}{September 22--26, 2025}{Prague, Czech Republic}
\acmBooktitle{Proceedings of the Nineteenth ACM Conference on Recommender Systems (RecSys '25), September 22--26, 2025, Prague, Czech Republic}\acmDOI{10.1145/3705328.3748073}
\acmISBN{979-8-4007-1364-4/2025/09}

\usepackage{etoolbox}
\makeatletter
\gdef\@copyrightpermission{
  \begin{minipage}{0.3\columnwidth}
   \href{https://creativecommons.org/licenses/by/4.0/}{\includegraphics[width=0.90\textwidth]{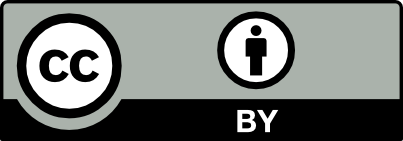}}
  \end{minipage}\hfill
  \begin{minipage}{0.7\columnwidth}
   \href{https://creativecommons.org/licenses/by/4.0/}{This work is licensed under a Creative Commons Attribution International 4.0 License.}
  \end{minipage}
  \vspace{5pt}
}

\begin{document}
\title{Hierarchical Graph Information Bottleneck for Multi-Behavior Recommendation}

\author{Hengyu Zhang}
\email{hyzhang@se.cuhk.edu.hk}
\authornote{Both authors contributed equally to this research.}
\affiliation{
  \institution{\text{The Chinese University of Hong Kong}}
  \city{Hong Kong SAR}
  \country{China}}

  \author{Chunxu Shen}
  \authornotemark[1]
\email{lineshen@tencent.com}
\affiliation{
  \institution{WeChat, Tencent Inc.}
  \city{Shenzhen}
  \country{China}}

  \author{Xiangguo Sun}
\email{xiangguosun@cuhk.edu.hk}
\affiliation{
  \institution{\text{The Chinese University of Hong Kong}}
  \city{Hong Kong SAR}
  \country{China}}

  \author{Jie Tan}
\email{jtan@se.cuhk.edu.hk }
\affiliation{
  \institution{\text{The Chinese University of Hong Kong}}
  \city{Hong Kong SAR}
  \country{China}}

  \author{Yanchao	Tan}
\email{yctan@fzu.edu.cn}
\affiliation{
  \institution{Fuzhou University}
  \city{Fuzhou}
  \country{China}}

  \author{Yu Rong}
\email{yu.rong@hotmail.com}
\affiliation{
  \institution{\text{The Chinese University of Hong Kong}}
  \city{Hong Kong SAR}
  \country{China}}

  \author{Hong Cheng}
  \authornote{Corresponding author.}
\email{hcheng@se.cuhk.edu.hk }
\affiliation{
  \institution{\text{The Chinese University of Hong Kong}}
  \city{Hong Kong SAR}
  \country{China}}

  \author{Lingling Yi}
  \authornotemark[2]
\email{chrisyi@tencent.com}
\affiliation{
  \institution{WeChat, Tencent Inc.}
  \city{Shenzhen}
  \country{China}}

\renewcommand{\shortauthors}{Hengyu Zhang et al.}

\begin{abstract}
  In real-world recommendation scenarios, users typically engage with platforms through multiple types of behavioral interactions. 
  Multi-behavior recommendation algorithms aim to leverage various auxiliary user behaviors to enhance prediction for target behaviors of primary interest (e.g., buy), thereby overcoming performance limitations caused by data sparsity in target behavior records. 
  Current state-of-the-art approaches typically employ hierarchical design following either cascading (e.g., view→cart→buy) or parallel (unified→behavior→specific components) paradigms, to capture behavioral relationships. However, these methods still face two critical challenges: (1) severe distribution disparities across behaviors, and (2) negative transfer effects caused by noise in auxiliary behaviors.
In this paper, we propose a novel model-agnostic Hierarchical Graph Information Bottleneck (HGIB) framework for multi-behavior recommendation to effectively address these challenges. Following information bottleneck principles, our framework optimizes the learning of compact yet sufficient representations that preserve essential information for target behavior prediction while eliminating task-irrelevant redundancies. To further mitigate interaction noise, we introduce a Graph Refinement Encoder (GRE) that dynamically prunes redundant edges through learnable edge dropout mechanisms. 
We conduct comprehensive experiments on three real-world public datasets, which demonstrate the superior effectiveness of our framework. Beyond these widely used datasets in the academic community, we further expand our evaluation on several real industrial scenarios and conduct an online A/B testing, showing again a significant improvement in multi-behavior recommendations.
The source code of our proposed HGIB is available at \url{https://github.com/zhy99426/HGIB}.
\end{abstract}



\maketitle

\section{Introduction}
\label{sec:intro}

In real-world online platforms, users typically engage with the platform through multiple types of interaction behaviors such as viewing, adding-to-cart, and purchasing. 
Therefore, the goal of multi-behavior recommendation tasks is to effectively leverage these multiple interaction behaviors to improve the prediction performance of the model for target behaviors (e.g., buy) of the primary focus of the platform.

To achieve this goal, researchers propose many advanced machine learning methods based on deep neural networks (DNNs)~\cite{dipn,matn}, graph convolutional networks (GCNs)~\cite{rgcn,mbgcn,gnmr,hago}, and attention mechanisms~\cite{matn,allinonefang, 10.1145/3477495.3532031,dualrec} to precisely model user preferences. Subsequent efforts explore incorporating multi-task learning (MTL) paradigms to utilize multi-behavior supervision signals~\cite{mgnn,ehcf,ghcf,mbgmn} and enhancing representation learning through self-supervised learning techniques like contrastive learning~\cite{cml,mbssl}.

\begin{figure}[]
    \centering
    \includegraphics[width=\linewidth]{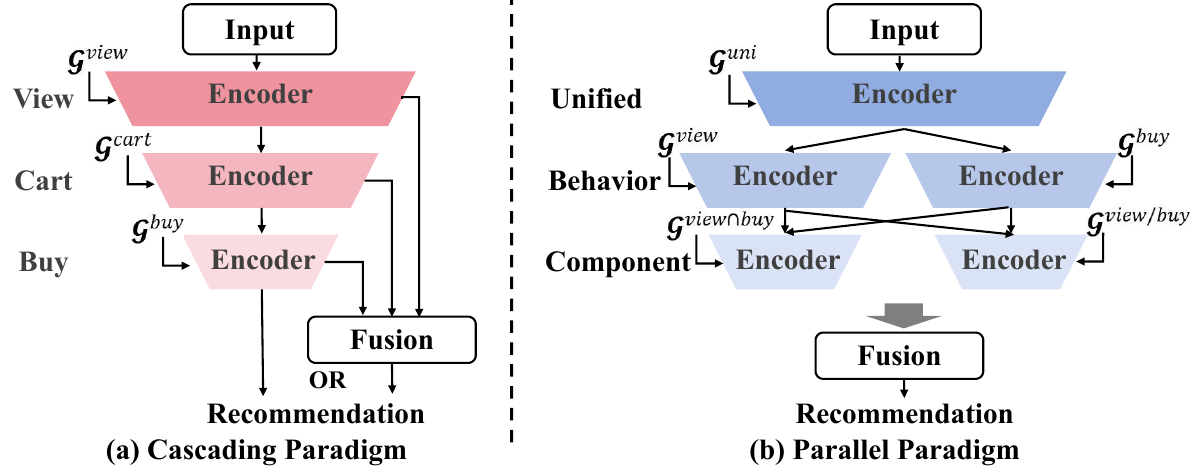}
    \caption{Illustration of the hierarchical model design in cascading and parallel paradigms.}
    \label{fig:intro}
\end{figure}

Recent state-of-the-art (SOTA) approaches predominantly adopt a hierarchical design philosophy to model the relationships between behaviors, primarily divided into two paradigms: cascading and parallel paradigms. 
Methods in the cascading paradigm~\cite{rgcn,mbcgcn,autodcs} are inspired by the natural cascading sequence of behaviors, such as view → cart → buy, modeling the relationships hierarchically through cascading encoders, as shown in Figure \ref{fig:intro} (a).
Since cascading relationships between behaviors are not strict, many other works adopt the parallel paradigm. Methods in the parallel paradigm~\cite{hpmr,mule} use parallel encoders for different behaviors and design modules to uncover relationships between them, such as decoupling "view" into "view and buy" and "view but not buy," as depicted in Figure \ref{fig:intro} (b).

\begin{figure}[!ht]
    \centering
    \begin{subfigure}{0.38\linewidth}
        \centering
        \includegraphics[width=\linewidth]{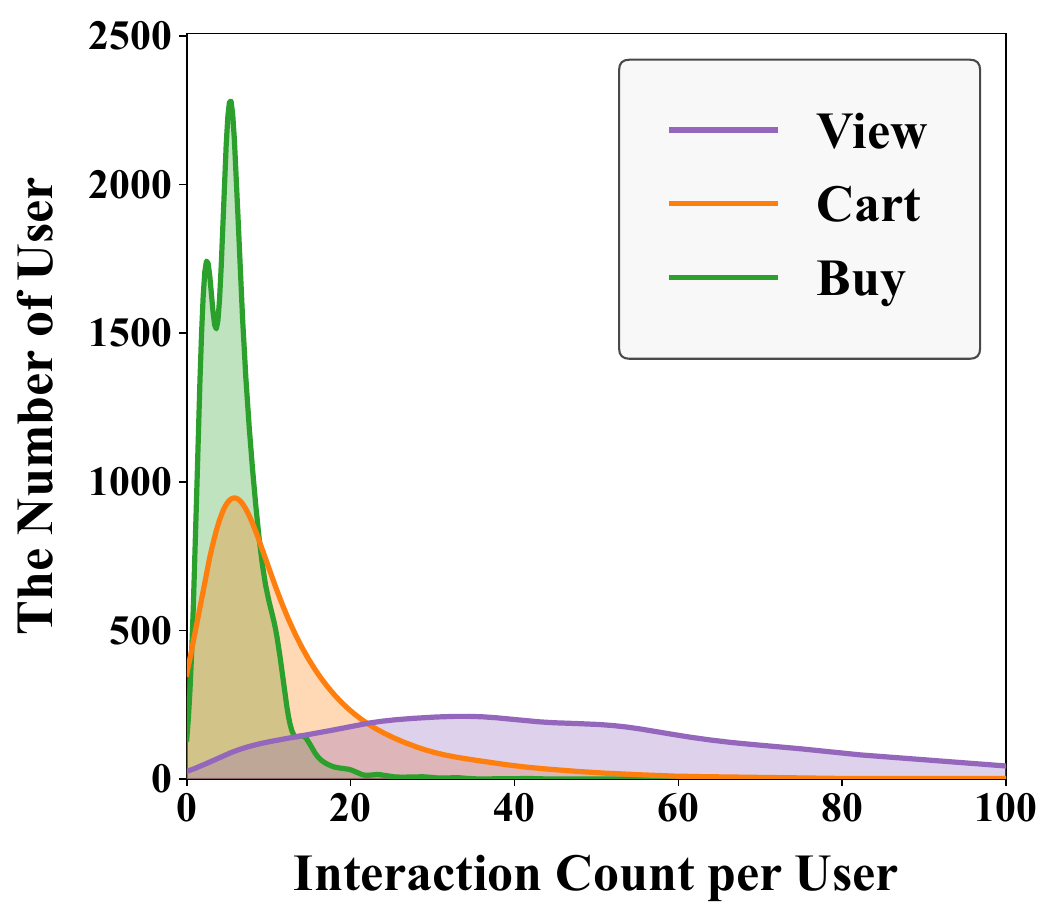} 
        \caption{}
    \end{subfigure}
    \begin{subfigure}{0.611\linewidth}
        \centering
        \includegraphics[width=\linewidth]{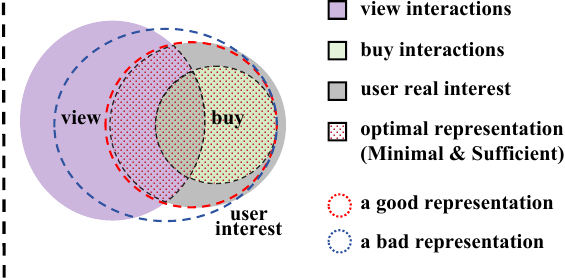} 
        \caption{}
    \end{subfigure}
    \caption{(a) Histogram of user numbers w.r.t interaction counts per user for different behaviors on the Taobao dataset. (b) Information diagram of learned representation, user real interest, and user-item interactions (e.g., view, buy).}
    \label{fig:intro2}
\end{figure}

Despite their success, the existing SOTA approaches still encounter critical challenges:

\begin{itemize}
    \item \textbf{Severe Distribution Disparities Across Behaviors.}
As illustrated in Figure \ref{fig:intro2} (a), there are substantial distributional differences among different behaviors. For instance, in e-commerce platforms, users may browse numerous items in a day but not make a purchase.
This imbalance introduces two key issues: On the one hand, auxiliary behaviors (e.g., view) typically exhibit substantially higher interaction frequencies than target behaviors (e.g., buy). So auxiliary behavior signals may introduce popularity bias into target behavior prediction due to their numerical dominance; On the other hand, the long-tail distribution of sparse target behaviors increases model overfitting risks. Figure \ref{fig:intro2}(b) further illustrates this issue: the red circle represents a good, sufficient, and compact user interest representation, while the blue circle is affected by bias from auxiliary behaviors and suffers from overfitting.

\item \textbf{Negative Transfer from Noisy Auxiliary Behaviors.}
Many auxiliary behaviors are implicit feedback and do not directly reflect users' real interests.
For example, a user browsing an item does not necessarily indicate a preference for it.
Additionally, feedback from accidental clicks is common.
These noises in auxiliary behaviors can negatively impact target behavior prediction, leading to a phenomenon known as negative transfer, which degrades model performance.
\end{itemize}

To overcome these challenges, we offer a simple yet effective framework for multi-behavior recommendation with \underline{\textbf{H}}ierarchical \textbf{\underline{G}}raph \textbf{\underline{I}}nformation \textbf{\underline{B}}ottleneck (\textbf{HGIB}). Specifically, our approach integrates information bottleneck principles into hierarchical model design, enforcing the multi-behavior model to retain the relevant information for the target task while minimizing irrelevant information.
In particular, we analyze the optimization objective of hierarchical models for multi-behavior recommendation from an information-theoretic perspective, and derive the optimization goal into the loss function by finding a lower bound of the maximized objective and employing HSIC~\cite{hsnorm} for approximations.
Furthermore, to explicitly remove noises from multi-behavior data, we propose the Graph Refinement Encoder (GRE) that prunes noisy interactions, thus further mitigating the negative transfer phenomenon from auxiliary behaviors. The main contributions of our work can be summarized as follows:

\begin{itemize}
    \item We analyze the optimization objective of hierarchical model design in multi-behavior recommendation from the perspective of the information bottleneck principle and present an innovative Hierarchical Graph Information Bottleneck (HGIB) framework. This framework effectively addresses the challenges of imbalanced multi-behavior data distribution and negative transfer.
    \item Our proposed Graph Refinement Encoder (GRE) explicitly removes noisy interactions from multi-behavior data, thereby alleviating negative transfer.
    \item Extensive experiments across three public datasets demonstrate the outstanding superiority and broad generality of our HGIB framework. Moreover, to further demonstrate the industrial application potential of our framework, we conduct comparison experiments on two large-scale industrial datasets with billions of interactions and perform a 15-day online A/B testing.

\end{itemize}

\section{Related Works}

\subsection{Multi-Behavior Recommendation}

The challenge of user interaction sparsity drives significant academic and industrial interest in multi-behavior recommendation systems, which aim to enhance target behavior prediction (e.g., buy) by leveraging auxiliary behavior interactions (e.g., view, cart). 
Early approaches primarily focus on extending single-behavior recommendation algorithms, exemplified by adopting matrix factorization into CMF~\cite{cmf}, the use of Monte Carlo sampling~\cite{10.1145/2566486.2567978}, or developing effective negative sampling strategies~\cite{ns1, ns2, mrbpr} for multi-behavior data.

The rise of deep learning techniques shifts research focus towards approaches based on neural networks. 
Researchers devote themselves to using advanced architectures, including deep neural networks (DNNs), graph convolutional networks (GCNs), and attention mechanisms to extract rich patterns from multi-behavior interactions. For example, DNN-based models like DIPN~\cite{dipn} and MATN~\cite{matn} employ attention mechanisms to model inter-behavior relationships for final representation aggregation. However, these architectures typically fail to capture high-order user-item interactions in graph structures, leading to GCN-based frameworks as the dominant paradigm. Representative works such as RGCN~\cite{rgcn}, MBGCN~\cite{mbgcn}, and GNMR~\cite{gnmr} construct unified interaction graphs to model global user preferences through graph convolution operations.
To better utilize auxiliary behavior signals, the multi-task learning framework is integrated into recommendation systems.
Some works like MGNN~\cite{mgnn}, EHCF~\cite{ehcf}, GHCF~\cite{ghcf}, CIGF~\cite{CIGF}, and MBGMN~\cite{mbgmn} achieve performance improvements through joint prediction of multiple behaviors. 
Subsequent efforts like CML~\cite{cml} and MBSSL~\cite{mbssl} seek to incorporate self-supervised learning techniques like contrastive learning to further enhance representation learning. More recently, HIRE~\cite{allinonefang} proposes a lightweight module to directly infer important heterogeneous interactions from data without relying on predefined patterns.

Recent advancements in cascading paradigms, as exemplified by CRGCN~\cite{crgcn} and MB-CGCN~\cite{mbcgcn}, leverage natural behavioral hierarchies (e.g., view→cart→purchase) to achieve superior prediction accuracy. 
DA-GCN~\cite{DA-GCN} extends the model of cascading relationships across multiple behaviors by constructing personalized directed acyclic behavior graphs.
However, these cascading models face challenges from distributional imbalances across behaviors, where biases and noise in auxiliary behaviors transfer to target behavior prediction, resulting in negative transfer issues. To address this limitation, PKEF~\cite{pkef} introduces a hybrid architecture combining parallel and cascading paradigms for behavior relationship modeling. The superiority of parallel paradigms is also demonstrated in MB-HGCN~\cite{mbhgcn} and HPMR~\cite{hpmr}. The state-of-the-art MULE~\cite{mule} model further demonstrates the superiority of parallel architectures through its unified→behavior→behavior component, achieving superior performance.

\subsection{Information Bottleneck for Recommendation}

The Information Bottleneck (IB) principle is successfully applied in machine learning to extract optimal intermediate representations that achieve sufficient compression while preserving predictive relevance. This principle has demonstrated significant success across various domains, including image classification, natural language understanding, and graph learning~\cite{yu2021graph}. In recent years, some research studies have tried to adapt the IB framework to recommendation systems, particularly in addressing challenges related to robust representation learning, denoising, debiasing, fairness, and explainability.
CGI~\cite{cgi} integrates the IB principle with contrastive learning frameworks through graph augmentation techniques like node and edge dropping to learn robust representations for recommendation tasks. DIB~\cite{dib, dib2} employs the IB framework to alleviate confounding bias in recommendation systems.  GBSR~\cite{gbsr} and IBMRec~\cite{IBMRec} leverage IB-based approaches for graph and feature denoising for social and multi-modal recommendation, respectively. FairIB~\cite{fairib} proposes a model-agnostic methodology that effectively balances recommendation accuracy with fairness considerations. These diverse applications underscore the versatility of the IB principle in addressing critical challenges across different recommendation tasks.
\section{Preliminaries}

\subsection{Problem Definition}
\label{sec:problem}

Given the user set $\mathcal{U} = \{u_1, u_2, \ldots, u_{|\mathcal{U}|}\}$ and item set $\mathcal{V} = \{v_1,$ $ v_2, \ldots, v_{|\mathcal{V}|}\}$, we formalize the multi-behavior recommendation task as follows. 
Let $\mathcal{B}$ denote the behavior set where $b_t \in \mathcal{B}$ represents the target behavior (e.g., buy) which is the primary concern of the platform, and $\mathcal{B}_{aux}=\mathcal{B} \setminus \{b_t\}$ are auxiliary behaviors (e.g., view, add-to-cart). 
Suppose $|\mathcal{B}|$ is the total number of behavior types, so there are $|\mathcal{B}|-1$ types of auxiliary behaviors in $\mathcal{B}_{aux}$. 
For each interaction that user $u$ has interacted with item $v$ under behavior $b \in \mathcal{B}$, a user-item pair $(u,v)$ is included in the edge set $\mathcal{E}^b$, which stores all the user-item interactions under behavior $b$. 
So we can treat the user-item interaction data under behavior $b$ as a form of bipartite graph, 
denoted as $\mathcal{G}^b=(\mathcal{U}\cup\mathcal{V}, \mathcal{E}^b)$.
The objective of multi-behavior recommendation is to utilize the data from auxiliary behaviors in $\mathcal{B}_{aux}$ sufficiently to assist in predicting the specific target behavior $b_t$.

\subsection{Information Bottleneck Principle}

The Information Bottleneck (IB) principle provides an information-theoretic framework for learning robust representations by discarding information that is not useful from the input. This framework formalizes the trade-off between preserving predictive information about the target task and compressing irrelevant variations in intermediate representation. 
Formally, given the input data $X$ and the target variable $Y$, IB aims to find an optimal representation $T$ which is sufficient and compact of input $X$ that maximally preserves information about the label $Y$ while minimizing redundancy, which can be formulated as:
\begin{equation}
    \max \underbrace{I(T; Y)}_{\text{Relevance}} - \beta\cdot \underbrace{I(X; T)}_{\text{Compression}},
    \label{eq:ib}
\end{equation}
where $I(T;Y)$ denotes the mutual information between the representation $T$ and the label $Y$, and $I(X;T)$ means the mutual information between the representation $T$ and the input $X$. $\beta > 0$ controls the trade-off of the parts of preservation and compression. 
Recently, the IB principle has become a powerful framework for robust representation learning in pattern recognition, natural language processing, and model explainability.

\section{Hierarchical Graph Information Bottleneck (HGIB)}
\label{sec:method}

\subsection{Overview}
\label{sec:overview}

In this section, we propose \underline{\textbf{H}}ierarchical \underline{\textbf{G}}raph \underline{\textbf{I}}nformation \underline{\textbf{B}}ottleneck (\textbf{HGIB}) framework, to integrate the information bottleneck principle into the hierarchical model design for multi-behavior recommendation. 
The overall objective of our proposed HGIB is to strategically preserve discriminative information relevant to target behavior prediction while eliminating irrelevant noise.

\subsubsection{Abstraction of Hierarchical Model}

To capture the complex interaction patterns between users and items, we map the user/item IDs in the given user set $\mathcal{U}$ and item set $\mathcal{V}$ to a learnable dense embedding, which can be formulated as:

\begin{equation}
    \boldsymbol{e}_u=E^T \cdot\boldsymbol{o}_u, \ \boldsymbol{e}_v=E^T \cdot\boldsymbol{o}_v,
\end{equation}

where $\boldsymbol{E}\in\mathbb{R}^{(|\mathcal{U}|+|\mathcal{V}|)\times d}$ denotes the embedding lookup table of user/item IDs, and $\boldsymbol{o}_u, \boldsymbol{o}_v\in \mathbb{R}^{|\mathcal{U}|+|\mathcal{V}|}$ mean the corresponding one-hot vector of user/item IDs. 

Denote the input embedding matrix $\boldsymbol{E}\in\mathbb{R}^{(|\mathcal{U}|+|\mathcal{V}|)\times d}$ as $\boldsymbol{E}^{0}$.
We can formulate the abstraction of the hierarchical model for recommendation as:

\begin{equation}
    \boldsymbol{E}^{l+1}=\operatorname{Encoder}_{l+1}(\boldsymbol{E}^{l}),
\end{equation}

where $\boldsymbol{E}^{l}$ means the output embedding matrix of the $l$-th encoder, and assuming there are $L$ encoders in total.

\subsubsection{Objective}

The objective of our proposed HGIB can be given by:

\begin{equation}
    \max\ \sum\limits_{l=1}^L\left[I(\boldsymbol{E}^{l};\boldsymbol{Y})-\beta\cdot I(\boldsymbol{E}^{l};\boldsymbol{E}^{l-1})\right],
    \label{eq:obj}
\end{equation}

where $\boldsymbol{Y}$ denotes the label matrix of the target behavior interactions. 
In Equation \ref{eq:obj}, the first term constrains that the embedding outputs from each encoder layer preserve task-relevant information crucial for target behavior recommendation, while the second term forces the encoder to progressively compress task-irrelevant components within the learned representations.

\begin{figure*}[!htp]
	\centering
	\includegraphics[width=\textwidth]{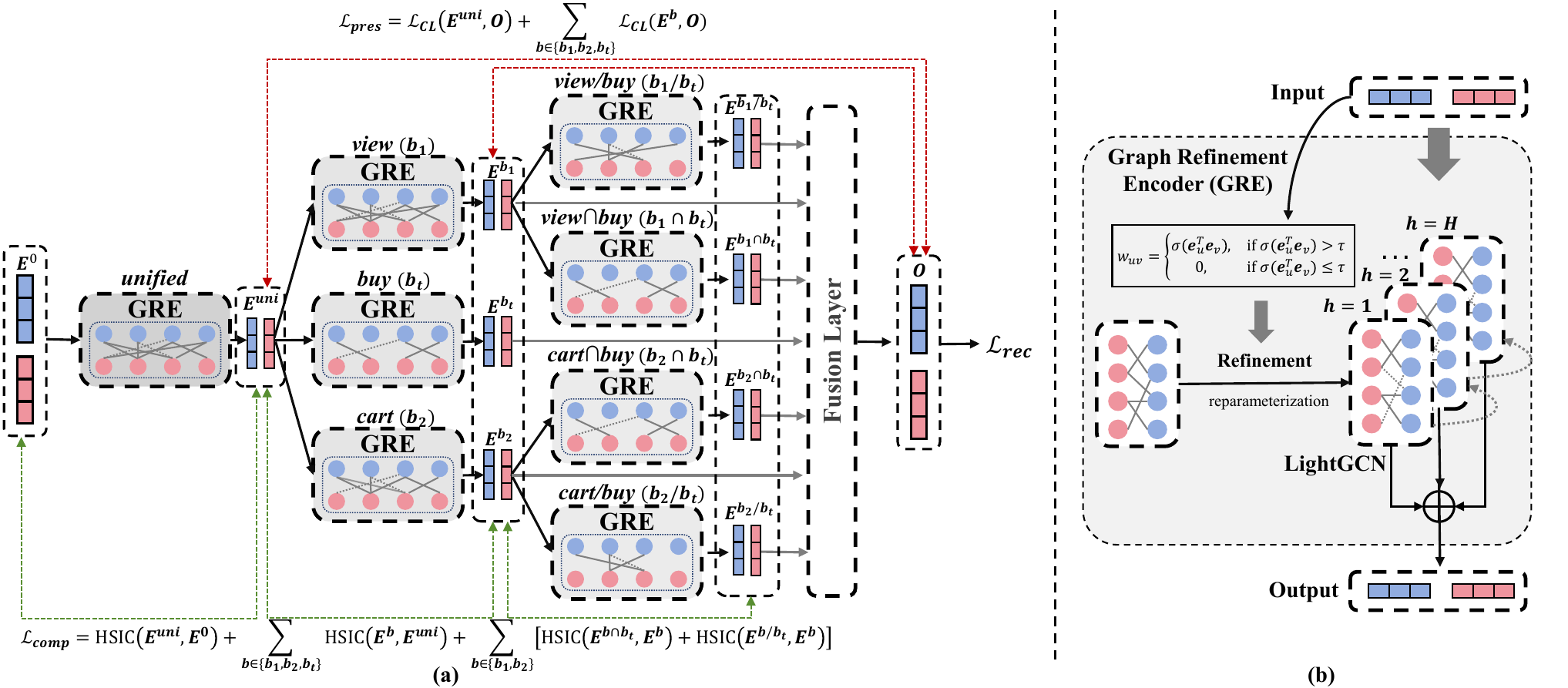}
	\caption{Illustration of the proposed HGIB, with three types of behavior as examples (view as $b_1$, cart as $b_2$, buy as $b_t$). (a) Instantiating of HGIB in the parallel paradigm, where the red dotted line represents $\mathcal{L}_{pres}$ for preservation and the green dotted line represents $\mathcal{L}_{comp}$ for compression. (b) Illustration of Graph Refinement Encoder (GRE), which denoises the noisy interaction by a learnable strategy.}

	\label{fig:model}
\end{figure*}

\subsection{Maximization of  $I(\boldsymbol{E}^{l};\boldsymbol{Y})$}
\label{sec:max}

Due to the intractability of directly computing mutual information, we strategically reformulate the optimization objective by deriving the lower bound of $I(\boldsymbol{E}^{l};\boldsymbol{Y})$ as follows:

\begin{equation}
\begin{aligned}
    \max \ I(\boldsymbol{E}^l;\boldsymbol{Y})&=I(\boldsymbol{E}^L;\boldsymbol{Y})+I(\boldsymbol{E}^l;\boldsymbol{Y}|\boldsymbol{E}^L)-I(\boldsymbol{E}^L;\boldsymbol{Y}|\boldsymbol{E}^l)\\
    &\geq I(\boldsymbol{E}^L;\boldsymbol{Y})-I(\boldsymbol{E}^L;\boldsymbol{Y}|\boldsymbol{E}^l)\\
    &=[H(\boldsymbol{Y})-H(\boldsymbol{Y}|\boldsymbol{E}^L)]-[H(\boldsymbol{E}^L|\boldsymbol{E}^l)-H(\boldsymbol{E}^L|\boldsymbol{E}^l,\boldsymbol{Y})]\\
    &\geq -H(\boldsymbol{Y}|\boldsymbol{E}^L) -H(\boldsymbol{E}^L|\boldsymbol{E}^l)\\
    &= -H(\boldsymbol{Y}|\boldsymbol{E}^L) -[H(\boldsymbol{E}^L)-I(\boldsymbol{E}^L;\boldsymbol{E}^l)]\\
    &= \underbrace{-H(\boldsymbol{Y}|\boldsymbol{E}^L)}_{\text{Recommendation}}\underbrace{+I(\boldsymbol{E}^L;\boldsymbol{E}^l)}_{\text{Preservation}}\underbrace{-H(\boldsymbol{E}^L)}_{\text{Regularity}}.
\end{aligned}
\label{eq:max}
\end{equation}

According to the derivation in ~\cite{infonce}, the second term $I(\boldsymbol{E}^L;\boldsymbol{E}^l)$ is lower bounded by minus InfoNCE ~\cite{infonce} loss as follows:
\begin{equation}
I(\boldsymbol{E}^L;\boldsymbol{E}^l)\geq \log(N)-\mathcal{L}_{CL}(\boldsymbol{E}^L,\boldsymbol{E}^l),
\label{eq:lower_bound}
\end{equation}
where $N$ denotes the number of samples including both positive and negative ones, and $\mathcal{L}_{CL}$ means the InfoNCE loss for contrastive learning. 

Moreover, the first term $-H(\boldsymbol{Y}|\boldsymbol{E}^L)$ is lower bounded by minus cross entropy loss, i.e., $-H(\boldsymbol{Y}|\boldsymbol{E}^L)\geq -\mathcal{L}_{CE}(\boldsymbol{E}^L, \boldsymbol{Y})$ where $\mathcal{L}_{CE}$ denotes the cross entropy loss, and the third term can be treated as the $L_2$-norm regularization of the trainable parameters of the model.
Thus, the maximum optimization of $I(\boldsymbol{E}^l;\boldsymbol{Y})$ is equivalent to the following training objective:

\begin{equation}
    \min \underbrace{\mathcal{L}_{CE}(\boldsymbol{E}^L, \boldsymbol{Y})}_{\mathcal{L}_{rec}}+\alpha\cdot\underbrace{\sum\nolimits_{l=1}^{L-1} \mathcal{L}_{CL}(\boldsymbol{E}^L,\boldsymbol{E}^l)}_{\mathcal{L}_{pres}} +\lambda\cdot \underbrace{||\Theta||^2_2}_{\mathcal{L}_{reg}},
\end{equation}

where $\mathcal{L}_{rec}, \mathcal{L}_{pres}, \mathcal{L}_{reg}$ are corresponding to recommendation, preservation, and regularization in Equation \ref{eq:max}, and $\Theta$ denotes the entire ensemble of trainable parameters in the model.

\subsection{Minimization of  $I(\boldsymbol{E}^{l};\boldsymbol{E}^{l-1})$}
\label{sec:min}

In the discussion of maximization of $I(\boldsymbol{E}^{l};\boldsymbol{Y})$, we find the lower bound of the mutual information $I(\boldsymbol{E}^L;\boldsymbol{E}^l)$ according to Equation \ref{eq:lower_bound}.
However, calculating the upper bound of mutual information is an intractable question.
To address this challenge, some prior studies~\cite{hsic,hsic2,gbsr} demonstrate that the Hilbert-Schmidt Independence Criterion (HSIC)~\cite{hsnorm} can serve as a great approximation in the minimization of mutual information. 
Given two random variables $ \boldsymbol{A}$ and $ \boldsymbol{B}$, HSIC measures dependence between them using kernel matrices. For $n$ observations $\{(a_i, b_i)\}_{i=1}^n$, define kernel matrices $\boldsymbol{K}^{A}$ (for $ \boldsymbol{A}$, $\boldsymbol{K}^{A}_{ij} = k^{A}(a_i, a_j)$) and $\boldsymbol{K}^{B}$ (for $ \boldsymbol{B}$, $\boldsymbol{K}^{B}_{ij} = k^{B}(b_i, b_j)$), centered by $\boldsymbol{H} = \boldsymbol{I} - \frac{1}{n}\mathbf{1}\mathbf{1}^\top$. The empirical HSIC is calculated as:

\begin{equation}
\text{HSIC}(A, B) = \frac{1}{(n-1)^2} \mathrm{tr}\left( \boldsymbol{K}^{A} \boldsymbol{H} \boldsymbol{K}^{B} \boldsymbol{H} \right),
\end{equation}

where $\mathrm{tr}(\cdot)$ denotes the matrix trace. HSIC is non-negative, and equals zero if and only if $A$ and $B$ are independent.

So the minimization optimization of $I(\boldsymbol{E}^{l};\boldsymbol{E}^{l-1})$ can be reformulated as follows:

\begin{equation}
    \min\ \mathcal{L}_{comp}=\sum_{l=1}^L \operatorname{HSIC}(\boldsymbol{E}^{l},\boldsymbol{E}^{l-1}),
\end{equation}

where $\mathcal{L}_{comp}$ correspond to the compression part in Equation \ref{eq:ib}.

\subsection{Denoising via Graph Refinement Encoder (GRE)}
\label{sec:gre}

In addition to the previously discussed robust representation learning guided by the information bottleneck principle, we further propose a Graph Refinement Encoder (GRE) that adopts a learnable mechanism to further explicitly eliminate noisy interactions in user-item graphs, as shown in Figure \ref{fig:model} (b). 
To achieve this purpose, we formulate the edge refinement strategy as follows:

\begin{equation}
    w_{uv}=\left\{
    \begin{aligned}
   \sigma(\boldsymbol{e}_u^T\boldsymbol{e}_v), & \quad \text{if } \sigma(\boldsymbol{e}_u^T\boldsymbol{e}_v) > \tau \\
    0, & \quad \text{if } \sigma(\boldsymbol{e}_u^T\boldsymbol{e}_v) \leq \tau
    \end{aligned}
    \right. 
\end{equation}

where $w_{uv}$ means the refined edge weights, and $\sigma(\cdot)$ denotes the sigmoid function and $\tau$ is the threshold to prune the noisy interactions.
Due to the non-differentiability of the strategy function, we apply the Gumbel-softmax~\cite{gumbel} reparameterization trick to enable end-to-end learning for the refinement strategy.
After refinement, GRE employs the LightGCN~\cite{lightgcn} for the graph aggregation process.

\subsection{Instantiation of HGIB}

In this section, we instantiate our proposed HGIB with a specific parallel paradigm backbone simplified from the strongest SOTA method MULE~\cite{mule}, as shown in Figure \ref{fig:model} (a). 

\subsubsection{Model Structure}
\label{sec:backbone}

Given the initial embedding $\boldsymbol{E}^0$ and user-item interaction graphs $\{\mathcal{G}^b\}_{b\in\{b_1,b_2,b_t\}}$, where $b_1, b_2, b_t$ represent view, cart, buy, the model structure of the HGIB instantiation can be summarized as follows:

\textbf{Unified Encoder:}
    \begin{equation}
    \boldsymbol{E}^{uni}=\operatorname{GRE}(\mathcal{G}^{uni}, \boldsymbol{E}^0),
    \end{equation}
    
    where $\mathcal{G}^{uni}=\bigcup_{b\in \{b_1, b_2,b_t\}} \{\mathcal{G}^{b}\}$, i.e., merging all the interactions from multiple behaviors, and $\operatorname{GRE}$ denotes the graph refinement encoder proposed in Section \ref{sec:gre}.

\textbf{Behavior-Specific Encoder:}
    \begin{equation}
        \boldsymbol{E}^{b}=\operatorname{GRE}(\mathcal{G}^{b}, \boldsymbol{E}^{uni}), \text{ for } b\in \{b_1, b_2,b_t\},
    \end{equation}
    
\textbf{Behavior-Component Encoder:}

    For $b\in \{b_1, b_2\}$,
    \begin{equation}
        \boldsymbol{E}^{b\cap b_t}=\operatorname{GRE}(\mathcal{G}^{b\cap b_t}, \boldsymbol{E}^{b}), \boldsymbol{E}^{b/b_t}=\operatorname{GRE}(\mathcal{G}^{b/b_t}, \boldsymbol{E}^{b}),
    \end{equation}
    
    where $\mathcal{G}^{b\cap b_t}=\mathcal{G}^{b}\bigcap\mathcal{G}^{b_t}$,i.e., the intersection of $\mathcal{G}^{b}$ and $\mathcal{G}^{b_t}$, indicates the user-item interaction graph of view->buy ($b_1\cap b_t$) and cart->buy ($b_2\cap b_t$), and $\mathcal{G}^{b/ b_t}=\mathcal{G}^{b}-\mathcal{G}^{b_t}$,i.e., the difference set of $\mathcal{G}^{b}$ and $\mathcal{G}^{b_t}$, indicates the user-item interaction graph of view-but-not-buy ($b_1/b_t$) and cart-but-not-buy ($b_2/b_t$).

\textbf{Fusion Layer:}
 Finally, we apply the widely-used Target Attention (TA)~\cite{din, dien, sim, dpn} to aggregate the final representation $\boldsymbol{O}$:
 
    \begin{equation}
    \begin{aligned}
        \boldsymbol{\hat E}&=\operatorname{TA}\left(\boldsymbol{E}^t,\{\boldsymbol{E}^{b}\}_{b\in \{b_1,b_2,b_t\}}\right),\\
        \boldsymbol{O}&=\operatorname{TA}\left(\boldsymbol{\hat E},\{\boldsymbol{E}^{b'}\}_{b'\in \{b\cap b_t,b/ b_t|b\in \{b_1, b_2\}\}}\right),
        \end{aligned}
    \end{equation}
    
    where $\operatorname{TA}(\boldsymbol{q},\boldsymbol{k})$ denotes apply the target attention to aggregate keys $\boldsymbol{k}$ with query $\boldsymbol{q}$.

\subsubsection{Optimization Objective}

Summarizing the discussion in Section \ref{sec:overview}-\ref{sec:min}, the objective function of HGIB instantiation can be formulated as follows:

\begin{equation}
    \min \mathcal{L}_{rec}+\alpha\cdot\mathcal{L}_{pres}+\beta\cdot\mathcal{L}_{comp}+\gamma\cdot\mathcal{L}_{reg},
\end{equation}

where $\mathcal{L}_{rec}$ is the cross-entropy loss for target behavior prediction and $\mathcal{L}_{reg}$ for $L_2$-norm regularization. The preservation loss $\mathcal{L}_{pres}$ and the compression loss $\mathcal{L}_{comp}$ can be given by:

\begin{equation}
\begin{aligned}
\mathcal{L}_{pres}&=\mathcal{L}_{CL}\left(\boldsymbol{E}^{uni},\boldsymbol{O}\right)+ \sum\limits_{b\in\{b_1,b_2,b_t\}}\mathcal{L}_{CL}\left(\boldsymbol{E}^b, \boldsymbol{O}\right),\\
    \mathcal{L}_{comp}&=\operatorname{HSIC}\left(\boldsymbol{E}^{uni}, \boldsymbol{E}^0\right)+\sum_{b \in\left\{b_1, b_2, b_t\right\}} \operatorname{HSIC}\left(\boldsymbol{E}^{b}, \boldsymbol{E}^{uni}\right)\\&+\sum_{b \in\left\{b_1, b_2\right\}}\left[\operatorname{HSIC}\left(\boldsymbol{E}^{b \cap b_t}, \boldsymbol{E}^{b}\right)+\operatorname{HSIC}\left(\boldsymbol{E}^{b / b_t}, \boldsymbol{E}^{b}\right)\right].
    \end{aligned} 
\end{equation}

\subsection{Further Analysis}

\subsubsection{Complexity Analysis}

Compared to the backbone model, the additional time complexity introduced by HGIB arises primarily from the calculation of the auxiliary loss $\mathcal{L}_{pres},\mathcal{L}_{comp}$ and Graph Refinement Encoder (GRE).
Because the time complexity of InfoNCE is $\mathcal{O}(B^2d)$ where $B$ denotes the batch size, the additional complexity from $\mathcal{L}_{pres}$ is $\mathcal{O}(B^2Ld)$ where $L$ is the number of encoders. The time complexity of  $\mathcal{L}_{comp}$ can be given by $\mathcal{O}(L)$ times the complexity of calculating HSIC, which results in $\mathcal{O}(B^3L+B^2Ld)$. And the calculation for GRE takes $\mathcal{O}(|\mathcal{E}|Ld)$ time. In summary, the additional time complexity introduced by HGIB can be given by $\mathcal{O}\left((B^2+|\mathcal{E}|)Ld+B^3L\right)$. 
Since the time complexity of commonly used backbone models (e.g., LightGCN) is already at least $\mathcal{O}(|\mathcal{E}|Ld)$ ($|\mathcal{E}|\gg B$), the proposed HGIB framework does not impose a significant computational overhead on the backbone network.

\subsubsection{Compatibility Analysis}

Our proposed HGIB is a model-agnostic framework for multi-behavior recommendation, which can be seamlessly compatible with other hierarchical multi-behavior methods. For applications to methods without hierarchical design, HGIB can be degenerated into the standard IB. 
Our proposed HGIB exhibits strong generalization across different hierarchical multi-behavior methods, as validated by the comprehensive compatibility analysis experiments in Section \ref{sec:compatibility}.

\section{Experiments}

To comprehensively evaluate the performance of our proposed HGIB, we conducted extensive experiments on three public real-world datasets to explore the following research questions:

\begin{itemize}
    \item \textbf{(RQ1)} How does HGIB perform compared to other state-of-the-art multi-behavior methods?
    \item \textbf{(RQ2)} How do the different components affect the prediction performance of HGIB, respectively?
    \item \textbf{(RQ3)} Can HGIB be compatible with other multi-behavior methods?
    \item \textbf{(RQ4)} How do different hyperparameters affect the prediction performance of HGIB, respectively?
    \item \textbf{(RQ5)} How does HGIB perform when applied to real industrial scenarios?
    \item \textbf{(RQ6)} Why does HGIB achieve better performance?
\end{itemize}

\subsection{Experiment Setting}

\subsubsection{Dataset Description}

To investigate the effectiveness of our proposed HGIB, we conduct comprehensive experiments on three popular public datasets for multi-behavior recommendation, including Taobao, Tmall, and Jdata datasets.
All these datasets are gathered from the top e-commerce platforms, such as taobao.com and jd.com in China.
The statistical information of the datasets is summarized in Table \ref{tab:dataset}.
Taobao dataset includes three types of behavior: \textit{view}, \textit{add-to-cart}, and \textit{buy}, while Tmall and Jdata datasets also have \textit{add-to-collect} behavior apart from these.
For all datasets, we treat \textit{buy} as the target behavior.
For preprocessing, we follow the settings of the previous works for these datasets.

\begin{table}[!ht]
\centering
\caption{Statistics of public datasets for the experiment.}
\label{tab:dataset}
\resizebox{\linewidth}{!}{
\begin{tabular}{ccccccc}
\toprule
\textbf{Dataset} & \textbf{\#Users} & \textbf{\#Items} & \textbf{\#Views} & \textbf{\#Collects} & \textbf{\#Carts} & \textbf{\#Buys} \\\midrule
\textbf{Taobao}  & 15,449           & 11,953           & 873,954          & -                   & 195,476          & 92,180         \\
\textbf{Tmall}   & 41,738           & 11,953           & 1,813,498        & 221,514             & 1,996            & 255,586         \\
\textbf{Jdata}   & 93,334           & 24,624           & 1,681,430        & 45,613              & 49,891           & 321,883        \\
\bottomrule
\end{tabular}}
\end{table}

\subsubsection{Evaluation Metrics}

For all our experiments, we assess the Top-$K$ recommendation performance of methods for predicting the target behavior (i.e., \textit{buy}) by two widely used metrics: HR@$K$ (Hit Ratio) and NDCG@$K$ (Normalized Discounted Cumulative Gain), where $K=10$ for our evaluation specifically.
 HR@$K$ is a recall-based metric that measures the average fraction of correct items among the Top-$K$ recommendations. 
 NDCG@$K$, on the other hand, evaluates the quality of the ranking in the Top-$K$ recommendations by considering the position of each item, assigning higher importance to items ranked higher in the list.
And we apply the full-ranking setting to sort the entire item set to get the Top-$K$ recommendation results.

\subsubsection{Baseline Methods}

To comprehensively validate the effectiveness of our proposed HGIB, we conduct comparisons with diverse baseline models, which can be categorized into three groups:  
    (1) \textbf{Single behavior methods}: MF-BPR~\cite{bpr}, NCF~\cite{ncf}, LightGCN~\cite{lightgcn};  
    (2) \textbf{Multi-behavior methods without hierarchical design}: RGCN~\cite{rgcn}, GNMR~\cite{gnmr}, NMTR~\cite{nmtr}, MBGCN~\cite{mbgcn};  
    (3) \textbf{Multi-behavior methods with hierarchical design}: CRGCN~\cite{crgcn}, MB-CGCN~\cite{mbcgcn}, HPMR~\cite{hpmr}, PKEF~\cite{pkef}, AutoDCS~\cite{autodcs}, COPF~\cite{copf}, MULE~\cite{mule}.

\subsubsection{Implement Details}

For a fair comparison, we strictly follow the data partitioning protocols and settings in prior works \cite{crgcn, mbcgcn, mule} for training-testing splitting.
For baseline methods, we adopted their official open-source implementations.
We configure the embedding dimension as $d=64$, and set the batch size to $1024$ for all implemented methods.

For our proposed HGIB, the source code is available \footnote{\url{https://github.com/zhy99426/HGIB}}. The hyperparameters of HGIB are set as follows: the coefficient $\alpha$ is set to $1.0$ for the Taobao and Tmall dataset and $0.5$ for the Jdata dataset; the coefficient $\beta=50$ and the threshold $\tau=0.05$ in GRE are set as the default for all datasets. And the regularization coefficient $\lambda$ is fixed to $0.1$.
For the model's optimization, we use the Adam optimizer with a learning rate of $5\times 10^{-4}$, training the model for a maximum of 100 epochs.

\subsection{Overall Performance \textbf{(RQ1)}}

The overall evaluation of the proposed HGIB against thirteen baseline models is summarized in Table \ref{tab:overall}. 

\begin{table}[!htp]
\caption{Overall evaluation results of our proposed HGIB and state-of-the-art baselines on three real-world multi-behavior datasets. 
Results show the best-performing method in bold and the runner-up underlined. "Rel Impr." indicates the relative improvements compared to the strongest baseline. An asterisk (*) denotes statistical significance (p < 0.05) when comparing HGIB to the strongest baseline results. }
\resizebox{\linewidth}{!}{
\begin{tabular}{c|cc|cc|cc}
\toprule\midrule
\multirow{2}{*}{\textbf{Method}} & \multicolumn{2}{c|}{\textbf{Taobao}} & \multicolumn{2}{c|}{\textbf{Tmall}} & \multicolumn{2}{c}{\textbf{Jdata}} \\ \cline{2-7} 
                                 & \textbf{HR}     & \textbf{NDCG}     & \textbf{HR}     & \textbf{NDCG}    & \textbf{HR}     & \textbf{NDCG}    \\\midrule
MF-BPR                           & 0.0076                                                    & 0.0036                                                      & 0.0230                                                    & 0.0207                                                      & 0.1850                                                    & 0.1238                                                      \\
NCF                            & 0.0236                                                    & 0.0128                                                      & 0.0124                                                    & 0.0062                                                      & 0.2090                                                    & 0.1410                                                      \\
LightGCN                         & 0.0411                                                    & 0.0240                                                      & 0.0393                                                    & 0.0209                                                      & 0.2252                                                    & 0.1436                                                      \\\midrule
RGCN                             & 0.0215                                                    & 0.0104                                                      & 0.0316                                                    & 0.0157                                                      & 0.2406                                                    & 0.1444                                                      \\
GNMR                             & 0.0368                                                    & 0.0216                                                      & 0.0393                                                    & 0.0193                                                      & 0.3068                                                    & 0.1581                                                      \\
NMTR                             & 0.0282                                                    & 0.0137                                                      & 0.0536                                                    & 0.0286                                                      & 0.3142                                                    & 0.1717                                                      \\
MBGCN                            & 0.0509                                                    & 0.0294                                                      & 0.0549                                                    & 0.0285                                                      & 0.2803                                                    & 0.1572                                                      \\\midrule
CRGCN                            & 0.0855                                                    & 0.0439                                                      & 0.0840                                                    & 0.0442                                                      & 0.5001                                                    & 0.2914                                                      \\
MB-CGCN                          & 0.1233                                                    & 0.0677                                                      & 0.0984                                                    & 0.0558                                                      & 0.4349                                                    & 0.2758                                                      \\
HPMR                             & 0.1104                                                    & 0.0599                                                      & 0.0956                                                    & 0.0515                                                      & 0.3260                                                    & 0.2029                                                      \\
PKEF                             & 0.1385                                                    & 0.0785                                                      & 0.1277                                                    & 0.0721                                                      & 0.4334                                                    & 0.2615                                                      \\
AutoDCS                          & 0.1522                                                    & 0.0813                                                      & 0.1432                                                    & 0.0743                                                      & \underline{0.6174}                                                    & \underline{0.4559}                                                      \\
COPF                          & 0.1552                                                    & 0.0838                                                      & 0.1755                                                    & 0.0967                                                     & 0.5338                                                   & 0.3692                                                      \\
MULE                             & \underline{0.1939}                                                    & \underline{0.1109}                                                      & \underline{0.2109}                                                    & \underline{0.1165}                                                      & 0.5820                                                    & 0.4147                                                      \\\midrule
\textbf{HGIB}                    & \textbf{0.2203}$^*$       & \textbf{0.1214}$^*$      & \textbf{0.2427}$^*$      & \textbf{0.1287}$^*$      & \textbf{0.6552}$^*$      & \textbf{0.4747}$^*$      \\
Rel Impr.               & 13.62\%      & 9.47\%      & 15.08\%     & 10.47\%     & 6.12\%      & 4.12\%      \\
\midrule
\bottomrule
\end{tabular}}
\label{tab:overall}
\end{table}

Drawing from these results, we summarize the following observations:

\begin{itemize}
    \item Leveraging multi-behavior data is essential, as single-behavior methods (e.g., LightGCN) consistently underperform against most of the multi-behavior methods.
    \item Multi-behavior methods with hierarchical designs consistently achieve superior performance compared to other approaches, regardless of whether they follow a cascading or parallel paradigm. By hierarchically modeling the relationships between multiple behaviors, these methods significantly enhance recommendation accuracy.
    
    \item  Due to the high conversion rates in the Jdata dataset (e.g., 43\% of "collect" behaviors and 57\% of "cart" behaviors convert to "buy"), the natural cascading sequence of behaviors is more pronounced. Thus, the cascading paradigm tends to outperform the parallel paradigm in this dataset.
    
    \item Our proposed HGIB framework achieves consistently superior performance across all datasets, demonstrating its exceptional effectiveness. This is attributed to HGIB’s use of the information bottleneck principle, which compels the model to retain information relevant to target behavior prediction while discarding irrelevant information. Furthermore, the GRE module filters out noisy interactions in the multi-behavior user-item interaction graphs, thereby mitigating negative transfer.
\end{itemize}

\subsection{Ablation Study \textbf{(RQ2)}}

As discussed in Section \ref{sec:method}, the main contribution of HGIB includes the introduction of the auxiliary loss $\mathcal{L}_{pres}$, $\mathcal{L}_{comp}$, and the GRE module.
To evaluate the effectiveness of these three components, we conduct an ablation study experiment by removing each component from the HGIB framework, resulting in three variant models: "w/o $\mathcal{L}_{pres}$", "w/o $\mathcal{L}_{comp}$" and "w/o GRE", where "w/o $\mathcal{L}_{pres}$" and "w/o $\mathcal{L}_{comp}$" are implemented by removing the corresponding loss, and w/o GRE is implemented by replacing Graph Refinement Encoder (GRE) with LightGCN ~\cite{lightgcn}.

\begin{table}[!ht]
\caption{Ablation study results of different components in HGIB. {\color{BlueGreen}Cyan} indicates performance decrease compared to HGIB, with deeper shades representing greater relative changes.}
\resizebox{\linewidth}{!}{
\begin{tabular}{lcccccc}
\toprule
\multirow{2}{*}{Method} & \multicolumn{2}{c}{Taobao} & \multicolumn{2}{c}{Tmall} & \multicolumn{2}{c}{Jdata} \\\cline{2-7}
                        & HR          & NDCG         & HR         & NDCG         & HR         & NDCG         \\\midrule
HGIB                    & \textbf{0.2203}       & \textbf{0.1214}      & \textbf{0.2427}      & \textbf{0.1287}      & \textbf{0.6552}      & \textbf{0.4747}      \\\midrule
w/o $\mathcal{L}_{pres}$       & \cellcolor{BlueGreen!15}{0.1870}       & \cellcolor{BlueGreen!14}{0.1034}      & \cellcolor{BlueGreen!26}{0.1799}      & \cellcolor{BlueGreen!23}{0.0980}      & \cellcolor{BlueGreen!6}{0.6172}      & \cellcolor{BlueGreen!4}{0.4551}      \\
w/o $\mathcal{L}_{comp}$        & \cellcolor{BlueGreen!1}{0.2188}       & \cellcolor{BlueGreen!1}{0.1204}      & \cellcolor{BlueGreen!4}{0.2322}      & \cellcolor{BlueGreen!5}{0.1222}      & \cellcolor{BlueGreen!3}{0.6397}      & \cellcolor{BlueGreen!5}{0.4534}     \\
w/o GRE         & \cellcolor{BlueGreen!4}{0.2110}       & \cellcolor{BlueGreen!4}{0.1167}      & \cellcolor{BlueGreen!9}{0.2200}      & \cellcolor{BlueGreen!9}{0.1164}      & \cellcolor{BlueGreen!4}{0.6319}      & \cellcolor{BlueGreen!5}{0.4491}     \\
\bottomrule
\end{tabular}}
\label{tab:ablation}
\end{table}

The results of the ablation study on three multi-behavior datasets are presented in Table \ref{tab:ablation}.
Comparing the results, we can draw the following observations:

\begin{itemize}
    \item The experiment results demonstrate that eliminating any key component results in a decline in model performance, thereby demonstrating the essential role and effectiveness of each component within the proposed HGIB framework. Specifically, both the preservation of task-relevant information and the compression of redundant information are necessary in the information bottleneck principle for robust representation learning. Furthermore, the explicit denoising implemented by GRE effectively mitigates the negative transfer problem caused by noise interference.

    \item The variant model "w/o $\mathcal{L}_{pres}$" exhibits the most significant performance degradation compared to HGIB, highlighting the critical importance of retaining task-relevant information in the information bottleneck framework. In multi-behavior recommendation scenarios, the $\mathcal{L}_{pres}$ objective requires the model to extract valuable knowledge specifically beneficial for the target behavior through representation learning. Building upon this foundation, the $\mathcal{L}_{comp}$  and GRE further compel the model to eliminate noise information irrelevant to the target behavior recommendation task. 
\end{itemize}

\subsection{Compatibility Analysis with Different Backbones \textbf{(RQ3)}}
\label{sec:compatibility}

Our proposed HGIB is a model-agnostic framework capable of being integrated with other multi-behavior models with hierarchical design.
To evaluate its adaptability, we implement the SOTA model of the cascading paradigm, AutoDCS~\cite{autodcs}, and of the parallel paradigm, MULE~\cite{mule}, as backbones to integrate into the HGIB framework.
We conduct compatibility analysis experiments through comparison of performance between original backbone models and their enhanced variants integrated into the HGIB framework, and the results are presented in Table \ref{tab:compatibility}. The ``Base'' in Table \ref{tab:compatibility} means the backbone model of HGIB introduced in Section \ref{sec:backbone}.

\begin{table}[!ht]
\caption{Compatibility analysis with different backbone models on three multi-behavior datasets. "Rel Impr." indicates the relative improvements compared to the corresponding backbones, and deeper {\color{RedOrange}orange} shades represent greater gains.}
\resizebox{\linewidth}{!}{
\begin{tabular}{ccccccc}
\toprule
\multirow{2}{*}{Method} & \multicolumn{2}{c}{Taobao} & \multicolumn{2}{c}{Tmall} & \multicolumn{2}{c}{Jdata} \\\cline{2-7}
                        & HR          & NDCG         & HR         & NDCG         & HR         & NDCG         \\\midrule
Base                    & 0.1679       & 0.0970      & 0.1655      & 0.0890      & 0.6063      & 0.4442      \\
HGIB                    & \cellcolor{RedOrange!27}{0.2203}       & \cellcolor{RedOrange!23}{0.1214}      & \cellcolor{RedOrange!40}{0.2427}      & \cellcolor{RedOrange!38}{0.1287}      & \cellcolor{RedOrange!8.5}{0.6552}      & \cellcolor{RedOrange!7}{0.4747}      \\
Rel Impr.               & 31.21\%      & 25.15\%     & 46.65\%     & 44.61\%     & 8.07\%      & 6.87\%      \\\midrule
AutoDCS                 & 0.1522       & 0.0813      & 0.1432      & 0.0743      & 0.6174      & 0.4559      \\
HGIB(AutoDCS)           & \cellcolor{RedOrange!10}{0.1674}       & \cellcolor{RedOrange!14}{0.0926}      & \cellcolor{RedOrange!10}{0.1578}      & \cellcolor{RedOrange!13}{0.0841}      & \cellcolor{RedOrange!2}{0.6198}      & \cellcolor{RedOrange!2}{0.4571}      \\
Rel Impr.               & 9.99\%       & 13.90\%     & 10.20\%     & 13.19\%     & 0.39\%      & 0.26\%      \\\midrule
MULE                    & 0.1939       & 0.1109      & 0.2109      & 0.1165      & 0.5820      & 0.4147      \\
HGIB(MULE)              & \cellcolor{RedOrange!12.6}{0.2183}       & \cellcolor{RedOrange!7.5}{0.1192}      & \cellcolor{RedOrange!13.5}{0.2395}      & \cellcolor{RedOrange!8.8}{0.1268}      & \cellcolor{RedOrange!8.7}{0.6326}      & \cellcolor{RedOrange!7.9}{0.4475}      \\
Rel Impr.               & 12.58\%      & 7.48\%      & 13.56\%     & 8.84\%      & 8.69\%      & 7.91\%   \\\midrule
\bottomrule
\end{tabular}}
\label{tab:compatibility}
\end{table}

As the experimental results in Table \ref{tab:compatibility} show, HGIB consistently achieves significant performance gains across different backbone models.
The performance improvement mainly stems from the fact that HGIB follows the information bottleneck principle enforces preservation of knowledge relevant to the prediction of the target behavior during hierarchical representation learning, while compressing noise and bias introduced by auxiliary behaviors.
It demonstrates the broad effectiveness and general superiority of our proposed HGIB framework for multi-behavior recommendations.

\subsection{Hyperparameter Analysis \textbf{(RQ4)}}

In this section, we will discuss the key hyperparameters $\alpha$ and $\beta$ in our proposed HGIB controlling the extent of preservation and compression in representation learning, respectively.
To evaluate how these hyperparameters influence HGIB's performance, we conduct a comparative study by varying their performance across public multi-behavior recommendation datasets. During experiments, while testing different configurations of $\alpha$ and $\beta$, all other model parameters are maintained at their default settings to isolate their individual effects.

\begin{figure}[!h]
\centering
\begin{subfigure}{0.327\linewidth}
    \centering
    \includegraphics[width=1.02\linewidth]{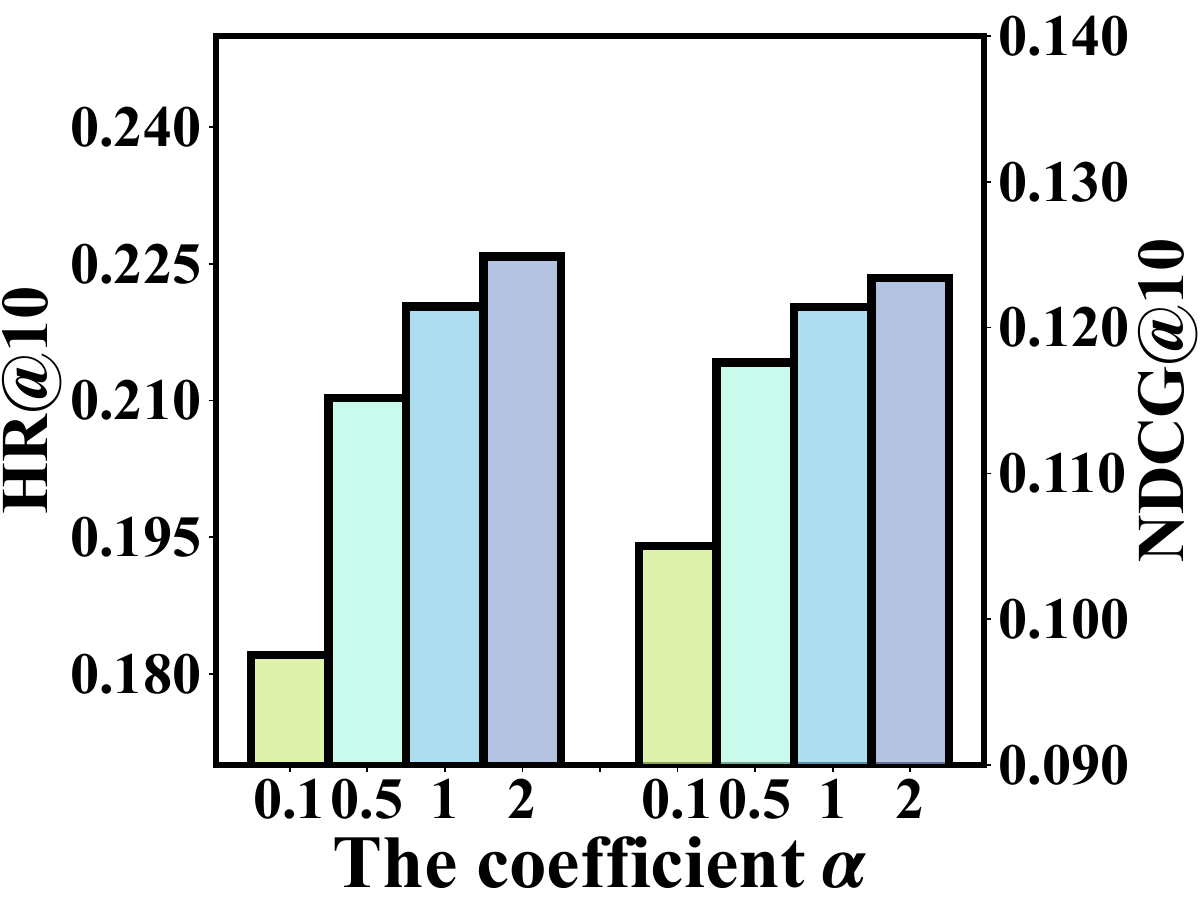} 
    \caption{Taobao}
\end{subfigure}
\begin{subfigure}{0.327\linewidth}
    \centering
    \includegraphics[width=1.02\linewidth]{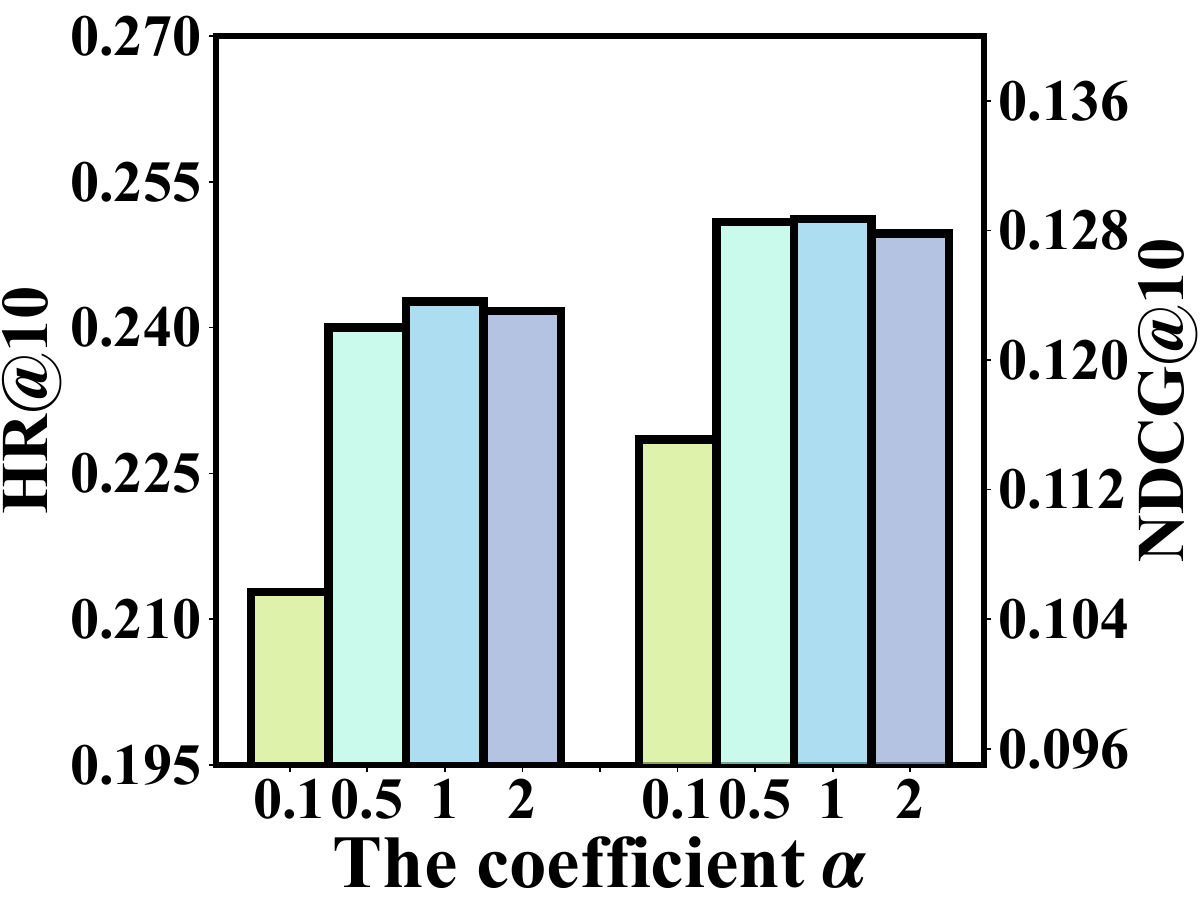} 
    \caption{Tmall}
\end{subfigure}
\begin{subfigure}{0.327\linewidth}
    \centering
    \includegraphics[width=1.02\linewidth]{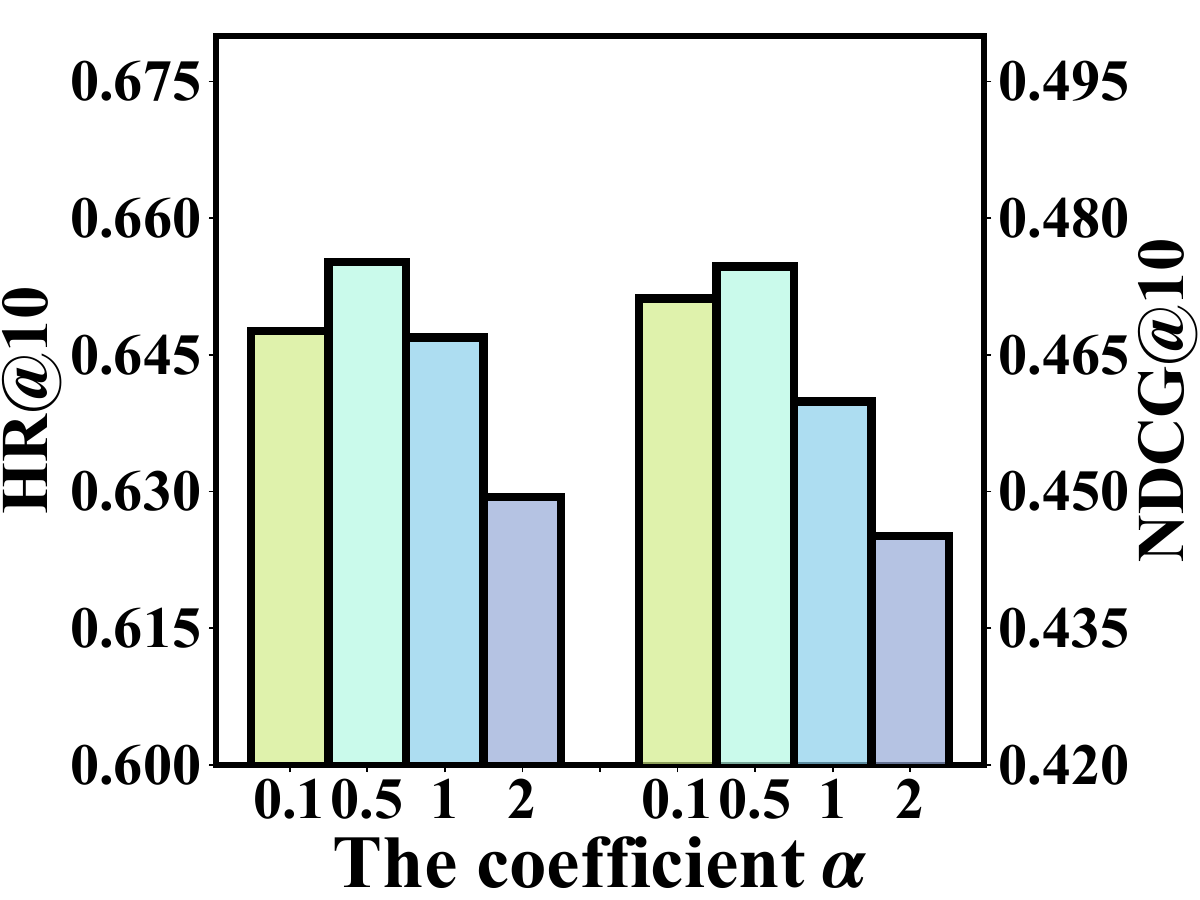} 
    \caption{Jdata}
\end{subfigure}

    \caption{Performance with the coefficient $\alpha$ for $\mathcal{L}_{pres}$.}
\label{fig:hyper1}
\end{figure}

\begin{figure}[!h]
\centering

\begin{subfigure}{0.327\linewidth}
    \centering
    \includegraphics[width=1.02\linewidth]{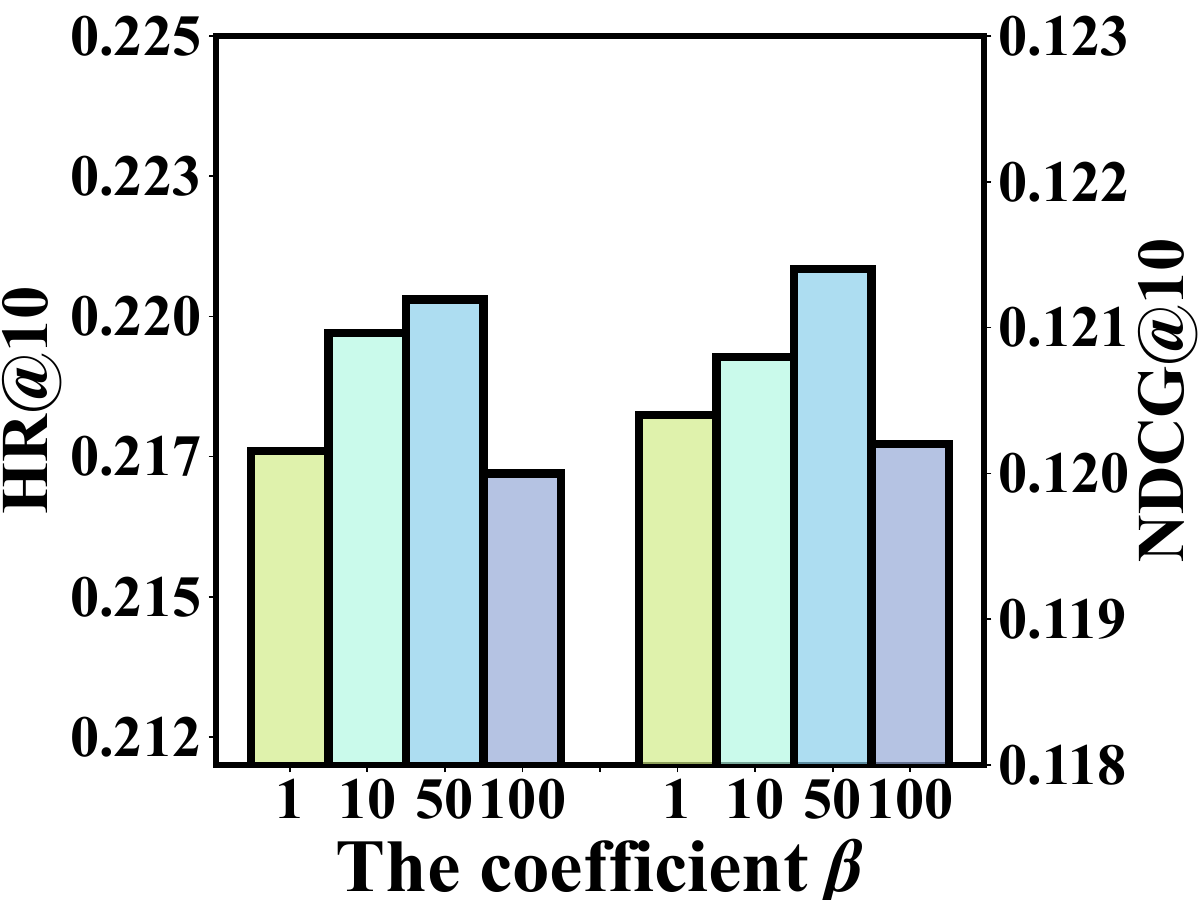} 
    \caption{Taobao}
\end{subfigure}
\begin{subfigure}{0.327\linewidth}
    \centering
    \includegraphics[width=1.02\linewidth]{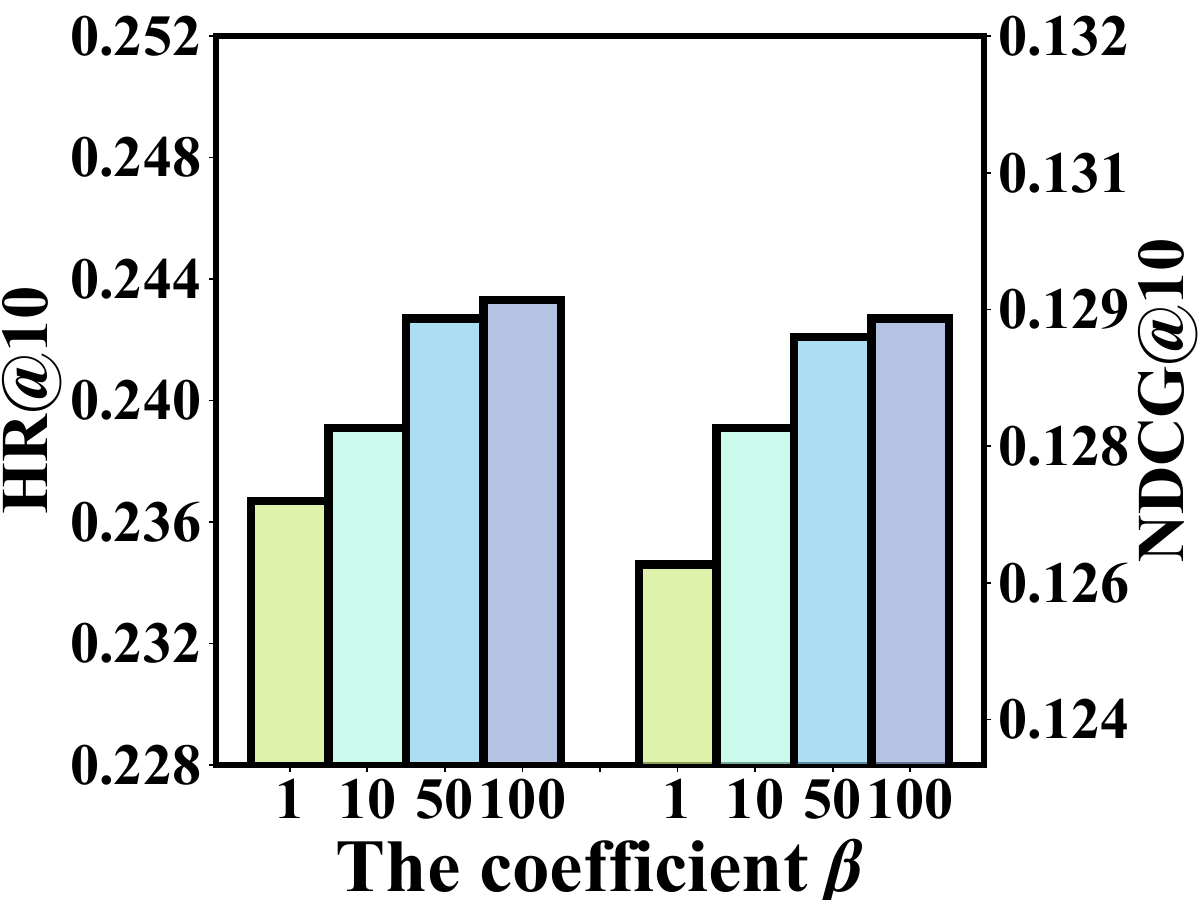} 
    \caption{Tmall}
\end{subfigure}
\begin{subfigure}{0.327\linewidth}
    \centering
    \includegraphics[width=1.02\linewidth]{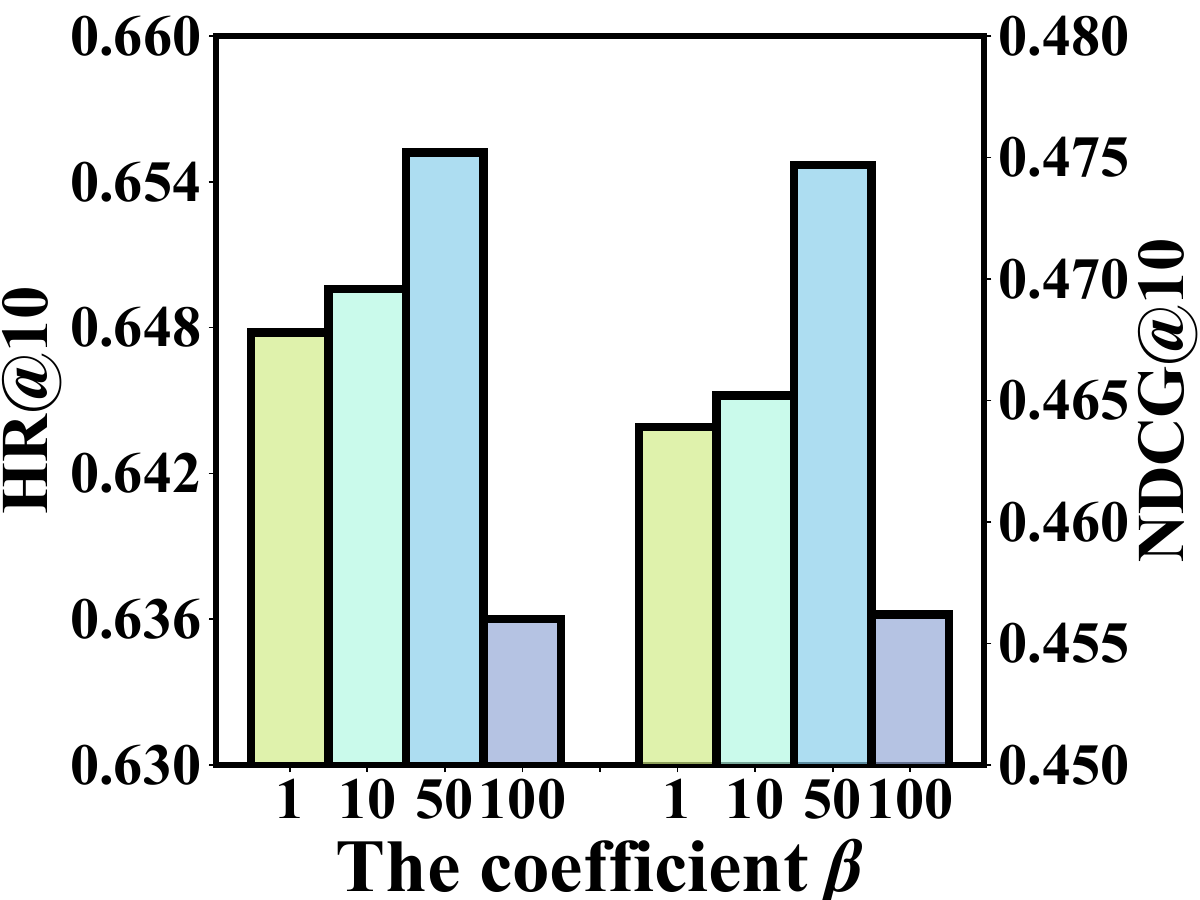} 
    \caption{Jdata}
\end{subfigure}

    \caption{Performance with the coefficient $\beta$ for $\mathcal{L}_{comp}$.}
\label{fig:hyper2}
\end{figure}

The comparative analysis of hyperparameters is illustrated in Figure \ref{fig:hyper1} and \ref{fig:hyper2}, leading to the following insights:

\begin{itemize}
    \item \textbf{The coefficient $\alpha$:} For Taobao and Tmall datasets, as $\alpha$ increases, the model is compelled to preserve task-relevant information in intermediate representations during the learning process, leading to gradual performance improvement. And performance gains plateau or even slightly decline when $\alpha$ exceeds to some extent.
    For the Jdata dataset, due to its inherent characteristics with a high conversion rate, the prediction task is relatively simple.
    Excessively large $\alpha$ values can disrupt the model's focus on the main recommendation task, resulting in peak performance occurring at $\alpha=0.5$.
    
    \item \textbf{The coefficient $\beta$:}
     Generally, HGIB achieves peak performance with a moderate $\beta$ value, so we set the default $\beta=50$ for all datasets.
    This occurs because overly small $\beta$ values fail to leverage the information bottleneck principle to compress task-irrelevant information; while excessively large $\beta$ values lead to over-compression, impairing the model’s prediction accuracy.

\end{itemize}

\subsection{Application on Industrial Scenarios \textbf{(RQ5)}}

Beyond the e-commerce scenarios covered by the three public datasets, our proposed HGIB framework demonstrates extensive applicability across diverse real-world industrial scenarios for multi-behavior recommendations.
To validate this, we conduct comparative experiments on two industrial datasets from \textbf{WeChat}, one of China's largest social platforms, regarding news recommendation and live-streaming recommendation scenarios.

\subsubsection{Experiment Description}

\begin{table}[!ht]
\centering
\caption{Statistics of industry datasets.}
\begin{tabular}{lrlr}
\toprule
\multicolumn{2}{c}{\textbf{News}} & \multicolumn{2}{c}{\textbf{Live Stream}} \\
\cmidrule(lr){1-2} \cmidrule(lr){3-4}
\textbf{Metric} & \textbf{Count} & \textbf{Metric} & \textbf{Count} \\\midrule
\#Users         & 26.81 million  & \#Users         & 32.25 million  \\
\#News          & 2.21 million   & \#Streamers     & 0.59 million   \\\midrule
\#Clicks        & 3.14 billion   & \#Clicks        & 2.64 billion   \\
\#Finished      & 1.46 billion   & \#Likes         & 0.15 billion   \\
\#Likes         & 29.64 million  & \#Comments      & 32.57 million   \\
\#Shares        & 42.79 million  & \#Gifts         & 16.60 million  \\
\bottomrule
\end{tabular}
\label{tab:ind_data}
\end{table}

The industrial datasets contain billions of interaction records from tens of millions of users across two scenarios: news recommendation (in a 15-day period) and live streaming recommendation (in a 90-day period). `Shares' is the target behavior for the news recommendation to promote social interaction and user stickiness, and `Gifts' is the target behavior for the live stream recommendation to increase the revenue of live streamers and the platform and promote the commercialization of the platform.

All data are subjected to rigorous privacy-preserving processing, and we follow the `leave-one-out' evaluation strategy for testing.
The detailed statistics of industrial datasets are shown in Table \ref{tab:ind_data}.

\begin{table}[!ht]
\caption{Performance comparison on industrial datasets.}
\begin{tabular}{ccccc}
\toprule
\multirow{2}{*}{Method} & \multicolumn{2}{c}{News} & \multicolumn{2}{c}{Live Stream} \\\cline{2-5}
                        & HR          & NDCG       & HR             & NDCG           \\\midrule
MULE                    & 0.1448      & 0.0701     & 0.1471         & 0.0712         \\
\textbf{HGIB}                    & \textbf{0.1663}      & \textbf{0.0825}     & \textbf{0.1552}         & \textbf{0.0764}         \\\midrule
Rel Impr.               & 14.85\%     & 17.69\%    & 5.51\%         & 7.30\%     \\\midrule 
\bottomrule
\end{tabular}
\label{tab:ind_result}
\end{table}

\subsubsection{Performance Comparison}

We compare our proposed HGIB with the strongest SOTA multi-behavior baseline, MULE~\cite{mule}, and evaluate performance using HR@10 and NDCG@10. The results, presented in Table \ref{tab:ind_result}, show that our method achieves relative improvements of $14.85\%$/$17.69\%$ in News Recommendation and $5.51\%$/$7.30\%$ in Live Stream Recommendation over MULE in terms of HR and NDCG. These results demonstrate that our proposed HGIB can effectively leverage the rich and diverse multi-behavior data of users for large-scale industrial recommendations.

\subsection{Online A/B Testing \textbf{(RQ5)}}

To evaluate our model's online performance in real industrial scenarios, we conducted a 15-day online A/B testing in the news recommendation ranking task of the WeChat platform from April 20, 2025, to May 5, 2025. WeChat Subscription Accounts provide daily news recommendation services for billions of users. We randomly served 2\% online traffic with our model, which contains over 20 million users, and another 2\% with the SOTA baseline model, MULE.

We used widely used metrics including \textbf{Click-Through Rate} (CTR) and \textbf{Social Sharing Rate} (SSR), to evaluate the online effectiveness.
The experimental results demonstrated statistically significant improvements of \textbf{0.32\%} in CTR (p<0.05) and \textbf{1.65\%} in SSR (p<0.01) over the SOTA baseline. The longitudinal validation demonstrates HGIB's effectiveness in enhancing user engagement on large-scale social platforms.

\subsection{Why It Works? \textbf{(RQ6)}}

In this section, we will discuss why our proposed HGIB achieves significant performance improvements compared to other baselines. To investigate this, we introduce the concept of information abundance from  \cite{multi_emb}, which measures the richness of information learned by embeddings. A low information abundance indicates that the embedding set tends to collapse into a low-rank matrix. 

\begin{definition}[Information Abundance]
Given an embedding matrix \( \boldsymbol{E} \in \mathbb{R}^{M \times d} \) where $M$ denotes the number of embeddings, and its singular value decomposition \( \boldsymbol{E} = \boldsymbol{U} \boldsymbol{\Sigma} \boldsymbol{V} = \sum_{k=1}^{d} \sigma_k \boldsymbol{u}_k \boldsymbol{v}_k^\top \). The information abundance of \( \boldsymbol{E} \) is defined as:
\[
\text{IA}(\boldsymbol{E}) = \frac{\|\boldsymbol{\sigma}\|_1}{\|\boldsymbol{\sigma}\|_\infty},
\]
i.e., the sum of all singular values normalized by the maximum singular value.
\end{definition}

\begin{figure}[!ht]
    \centering
    \includegraphics[width=\linewidth]{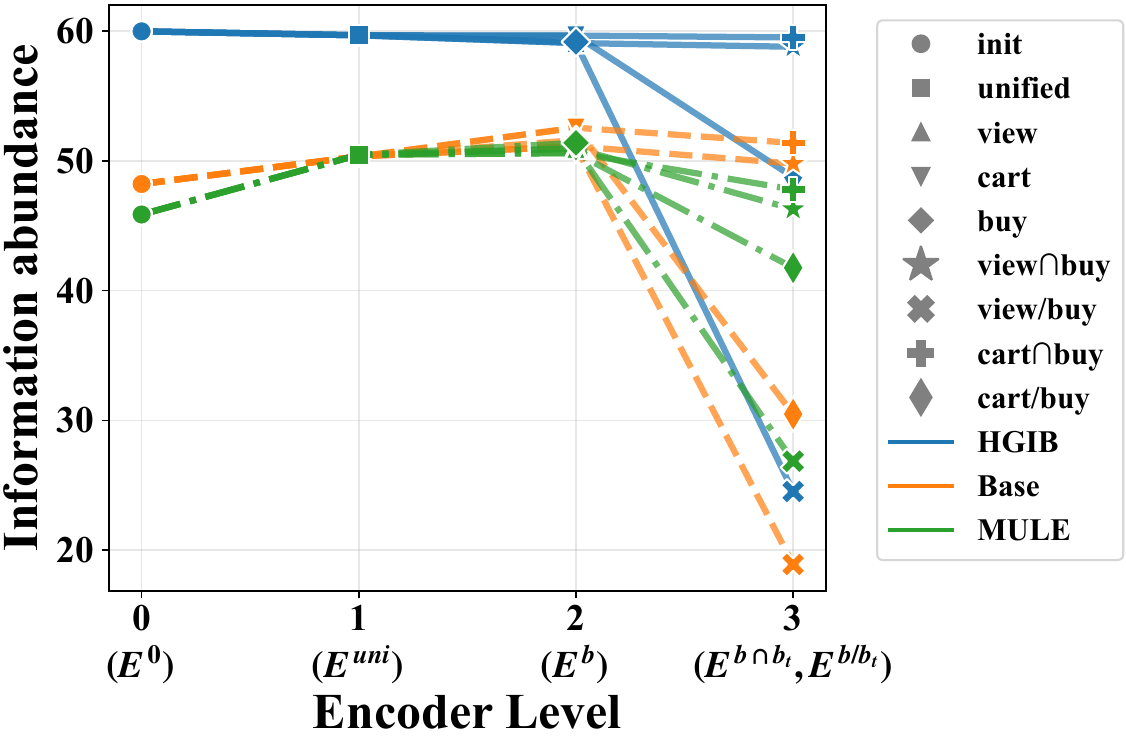}
    \caption{Comparison of information abundance of the embeddings in different models.}
    \label{fig:ia}
\end{figure}

We visualize the information abundance of embeddings output by each encoder on the Taobao dataset in our proposed HGIB, the backbone Base, and the baseline MULE, as shown in Figure \ref{fig:ia}.

In Figure \ref{fig:ia}, we can clearly observe that the embeddings learned by HGIB exhibit the highest information abundance. Additionally, the information abundance of embeddings in HGIB decreases with increasing encoder levels. This aligns with the essence of the hierarchical funnel design: the shallow unified graph contains the richest interaction data, including views, carts, and buys, while subsequent encoders model subgraphs that capture different behaviors and their relationships. As the encoder level increases, HGIB extracts task-relevant information while discarding other redundant information. In contrast, the shallow-encoder embeddings learned by the Base and MULE show insufficiently learned knowledge with low information abundance. This deviates from the principle of hierarchical design, limiting the learning of subsequent encoders and the overall performance of the model.

\section{Conclusions}

In this paper, we highlight critical challenges faced by the multi-behavior recommendation methods, including imbalanced behavior distributions and negative transfer caused by noisy interactions. To address these challenges, we proposed the Hierarchical Graph Information Bottleneck (HGIB) framework, which integrates information bottleneck principles into hierarchical multi-behavior modeling to retain task-relevant knowledge while compressing noise and irrelevant information. By optimizing an information-theoretic objective and employing the Graph Refinement Encoder (GRE) to prune noisy interactions explicitly, HGIB performs robust representation learning to mitigate overfitting and negative transfer effects. Extensive experiments on public and industrial datasets validate the superiority and generality of our approach, demonstrating its effectiveness in multi-behavior representation learning and its potential to enhance real-world recommendation systems.

\begin{acks}
This work is sponsored by Tencent WeChat Rhino-Bird Focused Research Program. Xiangguo Sun and Hong Cheng are supported by project \#MMT-p2-23 of the Shun Hing Institute of Advanced Engineering, The Chinese University of Hong Kong and by grant from the Research Grants Council of the Hong Kong Special Administrative Region, China (No. CUHK 14217622).
\end{acks}

\bibliographystyle{ACM-Reference-Format}
\bibliography{ref}


\end{document}